\documentclass[aps,prl,reprint,groupedaddress]{revtex4-1}

\usepackage{graphicx}
\usepackage{color}
\usepackage{ulem}
\begin{document}


\title{Momentum-space spectroscopy for advanced analysis of dielectric-loaded surface
plasmon polariton coupled and bent waveguides}

\author{K. Hassan} \author{A. Bouhelier}\email{alexandre.bouhelier@u-bourgogne.fr}\author{T. Bernardin}\author{G. Colas-des-Francs}
\author{J-C. Weeber}
\affiliation{Laboratoire Interdisciplinaire Carnot de Bourgogne, UMR 6303 CNRS-Universit{\'{e}} de Bourgogne,\\
 9 avenue A. Savary, BP 47870, F-21078 Dijon, France}
\author{R. Espiau de Lamaestre}
\affiliation{CEA LETI, MINATEC Campus, 17 rue des Martyrs 38054
Grenoble cedex 9, France}
\author{A. Dereux}
\affiliation{Laboratoire Interdisciplinaire Carnot de Bourgogne, UMR 6303 CNRS-Universit{\'{e}} de Bourgogne,\\
 9 avenue A. Savary, BP 47870, F-21078 Dijon, France}
\date{\today}

\date{\today}

\begin{abstract}
We perform advanced radiation leakage microscopy of routing dielectric-loaded plasmonic waveguiding structures.
By direct plane imaging and momentum-space
spectroscopy, we analyze the energy transfer between coupled
waveguides as a function of gap distance and reveal the momentum
distribution of curved geometries. Specifically, we observed a clear
degeneracy lift of the effective indices for strongly interacting
waveguides in agreement with coupled-mode theory.
We use momentum-space representations to discuss the effect of curvature on dielectric-loaded waveguides. The
experimental images are successfully reproduced by a numerical and an analytical model of the mode propagating in a curved plasmonic waveguide.
\end{abstract}


\maketitle


\section{Introduction}

Confinement and propagation of surface plasmons in a metal circuitry
have received considerable interest for their capability to
transport data with a large bandwidth in compact structures and
devices. Among the different geometries capable of routing the
flow of surface plasmon, dielectric-loaded surface plasmon polariton
waveguides (DLSPPWs)~\cite{krenn05OL,Holmgaard08,grandidier08} have
recently emerged as a potential plasmonic architecture that can be
integrated seamlessly with current silicon-on-insulator (SOI)
photonic circuits~\cite{briggs10NL,pleros12} and can sustain transfer
of information at high date rates~\cite{Papaioannou11}. A DLSPPW is
made of a rectangular dielectric rib deposited on a metal film or
strip~\cite{Grandidier10}. The surface plasmon is confined in the
dielectric layer and typical cross-sections required for an optimum
confinement of the mode compare well with state-of-the-art SOI
waveguides operating in the telecom bands. Despite dramatically
higher losses, the advantage of such plasmonic platform is that the
optical index of dielectric material used to confined the mode can
be externally controlled to realize active DLSPPW-based
devices~\cite{grandidier09,Gosciniak10,randhawa12,Perron:11,Krasavin:11,hassan:11}.

Understanding device performance such as transmission loss and
surface plasmon mode profile greatly contributed to the development
of DLSPPWs. Near-field optical microscopy and far-field
leakage radiation microscopy (LRM)~\cite{hecht96PRL} are
instrumental for visualizing the confinement and propagation details
of surface plasmons supported by this
geometry~\cite{holmgaard08PRB,steinberger07,grandidier08}. For thin
metal films, LRM is an especially useful characterization tool. It
provides a diffraction-limited snapshot of the mode profile in the
structure, and, by conoscopic imaging enables to extract the
wave-vector distribution. Effective indices of the different modes
and interactions developing in a given structure can thus be readily
determined~\cite{grandidier08,krishnan10,Berthelot11,Regan12}.

In this work, we performed an in-depth analysis of leakage radiation
images obtained for two different DLSPPW-based routing
structures: linear couplers and bent waveguides. By simultaneously
imaging the conjugated aperture and field planes of the microscope
we unambiguously quantify the degeneracy lift occurring for strongly
interacting DLSPPWs and directly visualize the symmetry of the coupled modes.
Furthermore, we measured the wave-vector distribution of
90$^\circ$-curved DLSPPWs and show its evolution with bend radius.
The experimental images are compared to numerical and analytical calculations.

\section{Experimental section}
\subsection{DLSPPW fabrication}
\begin{figure}
\includegraphics[width=8.6cm]{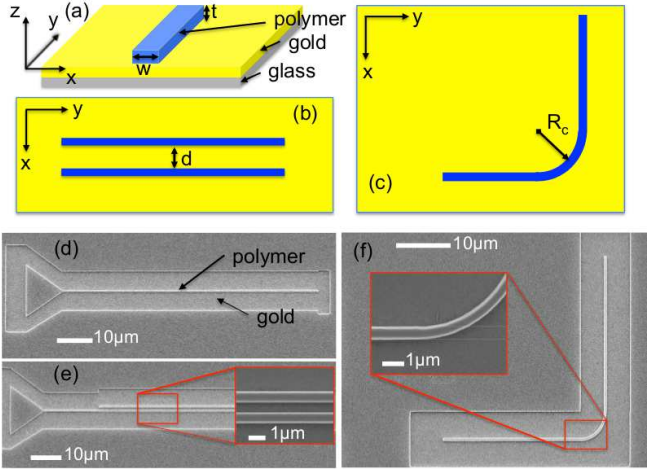}
\caption{(a) Schematic view of a basic DLSPPW with a width $w$ of
600~nm and thickness $t$=560~nm. (b) Schematic view of two coupled
waveguides separated by a gap $d$. (c) Schematic view of a curved waveguide with a bend radius $R_c$. (d), (e) and (f) are
scanning electron micrographs of typical devices corresponding to
the configurations (a), (b), and (c), respectively.}\label{fp_Fig1}
\end{figure}

The waveguides considered in this work are depicted schematically in
Figs.~\ref{fp_Fig1} (a) to (c). A Poly(methylmethacrylate) (PMMA)
ridge is defined by electron-beam lithography onto a 65 nm-thick gold
film evaporated on a glass substrate. The thickness of all DLSPPWs
fabricated in this study is fixed at $t$=560~nm and the width at
$w$=600~nm. For such dimensions, the DLSPPW structures shown in
Fig.~\ref{fp_Fig1} are single-mode at telecom wavelength~\cite{holmgaard07}. Scanning
electron micrograph of the routing elements are shown in
Figs.~\ref{fp_Fig1} (d) to (f).

\subsection{Characterization setup}
\begin{figure}
\includegraphics[width=8.6cm]{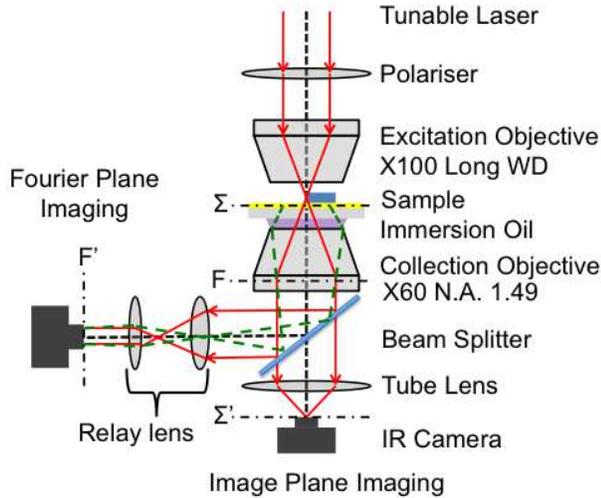}
\caption{Schematic of the leakage radiation microscope used in
this study. Launching of the plasmon mode in the waveguides is
obtained by focusing a laser operating at $\lambda$=1510~nm with a
long working distance objective. Surface plasmon leakages are
collected by an oil immersion objective and are imaged on two
cameras placed at the conjugated image (field) plane and Fourier
(aperture) plane, respectively.} \label{fp_Fig2}
\end{figure}
Optical characterization is performed by using a
home-made leakage radiation microscope (LRM). The principle of this
method is only recalled here, more detailed can be found in the
review by Drezet \textit{et al.}\cite{drezet08review}. The schematic
view of our experimental set-up is illustrated in
Fig.~\ref{fp_Fig2}. An incident tunable laser beam (fixed at $\lambda$=1510~nm hereafter) is focused by a 100 $\times$ microscope objective on the extremity of a DLSPP waveguide. The sharp discontinuity defined by the polymer structure acts a local scatterer and produces a spread of wave-vectors, some of which resonant with the surface plasmon modes supported by the geometry. By controlling the incident polarization parallel to the longitudinal axis of the waveguide~\cite{Frisbie10}, the DLSPPW mode can thus be readily excited.  Leakage radiation microscopy (LRM) provides a far-field imaging technique to directly visualize surface plasmon propagation and investigate its fundamental properties~\cite{hecht96PRL,bouhelier99}. This method relies on the collection of radiation losses occurring in the waveguide during propagation. These losses are emitted in the substrate at an angle phase-matched with the in-plane wave-vector of the SPP mode. In our LRM, the plasmon radiation losses are collected by an oil-immersion objective with a numerical aperture (N.A.) of 1.49. A tube lens focuses the leakages in an image plane (IP) conjugated with the object plane where an InGaAs infrared camera is recording the two-dimensional intensity distribution.  Images recorded at this plane provide direct information about the propagation of the surface plasmon mode developing in the DLSPPW. To complete the analysis, we have also used the Fourier transforming property of the objective lens to access the wave-vector distribution of the emitted light. The angular distribution of the rays radiated in the substrate and collected by the objective lens is transformed to a lateral distribution at the objective back focal plane. Quantitative measurement of the complex surface plasmon wave-vector, or equivalently its complex effective index, consists at measuring the radial distance of the rays with respect to the optical axis. Access to momentum space was done by inserting a beam splitter in the optical path and a set of Fourier transforming relay lens. The lenses are forming an image conjugated with back focal plane of the objective. This Fourier plane (FP) contains the two-dimensional wave-vector distribution of the leakage radiation and momentum-space spectroscopy can thus be readily performed.
\subsection{Calibrating leakage radiation images}
\begin{figure}
\includegraphics[width=8.6cm]{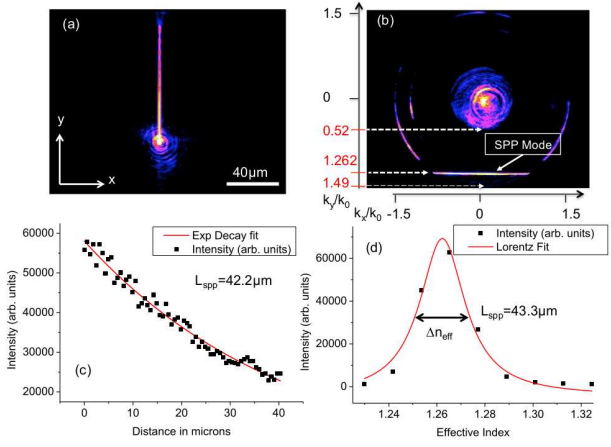}
\caption{(a) Intensity distribution recorded at the image plane of a
surface plasmon mode propagating along a straight DLSPPW. The
excitation spot is readily visible at the lower portion of the
waveguide. (b) Corresponding wave-vector distribution. The central
disk represents to the numerical aperture of the illumination
objective (0.52). The DLSPP mode is recognized as a bright line at
constant $k_y/k_x$ (arrow). (c) Exponential fit of the experimental decaying
plasmon intensity along the waveguide leading to a
$L_{spp}$=42.2~$\mu$m. (d) Lorentzian fit of the FP plasmon mode
signature centered at $\beta^{\prime}/k_o$=1.262. The full with at half maximum 
$\Delta n_{eff}$ is inversely proportional to $L_{spp}$.} \label{fp_Fig3}
\end{figure}

Figures~\ref{fp_Fig3}(a) and (b) show images of the leakage
radiation intensity of a single-mode DLSPPW recorded in the
conjugated image and Fourier planes, respectively. The plasmon mode
is launched at the bottom of the structure and propagates up the
waveguide with an exponentially decaying intensity. The image of
Fig.~\ref{fp_Fig3}(b) is a direct measurement of the wave-vector content
of the intensity distribution shown in Fig.~\ref{fp_Fig3}(a). For
both FP and IP images, a calibration was performed prior to any data
extraction. For calibrating IP images, we used the known waveguide
length as a standard. With the system magnification one pixel
represents $\simeq$0.6~$\mu$m. For FP images, the $N.A.$ of the
objective was used to calibrate $k_{x}$ and $k_{y}$ axes. The
largest ring in Fig.~\ref{fp_Fig3}(b) represents the $N.A.$=1.49
specified by the manufacturer. The central disk is the numerical
aperture of the $\times 100$ excitation objective at $N.A.$=0.52. One
pixel provides a $\Delta N.A.\simeq$ 0.012.

The DLSPPW mode propagating along this straight waveguide presented in
Fig.~\ref{fp_Fig3} can be defined by two parameters: its effective
index $n_{eff}$ and propagation length $L_{spp}$. $n_{eff}$ is
expressed by the phase constant $\beta^{\prime}$ and reads
$n_{eff}=\beta^{\prime}/k_o$ where $k_o=2\pi/\lambda$. The
propagation length $L_{spp}=(2\beta^{\prime\prime})^{-1}$ where
$\beta^{\prime\prime}$ is the attenuation constant of the plasmon
mode. A complex propagation constant is then evaluated from these two
constants: $\beta=\beta^{\prime}+i \beta^{\prime\prime}$.

$L_{spp}$ can be directly extracted from direct-space image by
fitting an exponential decay of the intensity along the waveguide
$I=I_0 \exp(-y/L_{spp})$. Here a $L_{spp}$=42.2~$\mu$m is determined
from the fit to the experimental data illustrated in
Fig.~\ref{fp_Fig3}(c). Experimentally, the FP image displayed in
Fig.~\ref{fp_Fig3}(b) contains more information than a direct space
image since the real part and imaginary part of the effective index
can be directly measured. The signature of the mode is represented
by a single line at a constant $n_{eff}$=$k_y/k_o$=1.262$\pm$0.006.
The intensity measured along $k_y/k_o$ at
$k_x$=0 is related to the surface plasmon through the following
formula~\cite{grandidier09}: $I(k_x,k_y)\propto
|\widetilde{H}_0(k_x)|^2/[(k_{y}-\beta^{\prime})^2+(1/2L_{spp})^2]$.
$\widetilde{H}_0(k_x)$ is the $k_x$-Fourier
transform of the guided magnetic field at the objective focal point.
The imaginary part
of the effective index is also estimated precisely through a Lorentzian
fit. The width $\Delta n_{eff}$ of the $\beta^{\prime}/k_o$ line is a measure of the losses
experienced by the plasmon mode and is thus inversely proportional
to its propagation length $L_{spp}$~\cite{Raether88}. We obtain from the FP image a
$L_{spp}$=43.3~$\mu$m in fairly good agreement with $L_{spp}$=42.2~$\mu$m measured from direct-space analysis.

\section{Momentum-space spectroscopy of linear DLSPPW couplers}

\begin{figure}
\includegraphics[width=8.6cm]{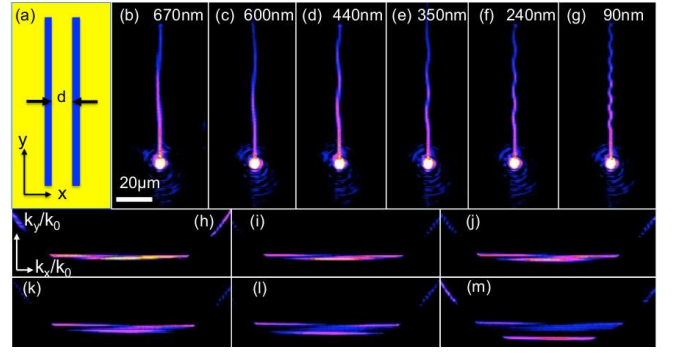}
\caption{(a) Schematic of a linear DLSPPW coupler. (b) to (g) are
direct-space images of the intensity distribution in linear DLSPPW
couplers with edge-to-edge distance $d$ equals to 670~nm, 600~nm,
440~nm, 350~nm, 240~nm and 90~nm respectively. (h) to (m) are the
corresponding wave-vector distributions unambiguously demonstrating
the degeneracy lift of the coupled waveguides for small values of $d$. The images were cropped to display only the lower part of the Fourier plane.} \label{fp_Fig4}
\end{figure}

We now demonstrate the added-value of performing momentum-space
leakage radiation spectroscopy of a classical integrated plasmonic
routing device: a linear DLSPPW
coupler~\cite{steinberger07,krasavin08,chen09,colas09}. The linear
coupler geometry illustrated in Fig.~\ref{fp_Fig4}(a) consists of
two parallel and identical DLSPP waveguides separated by an
edge-to-edge distance $d$ varying from 670~nm to 90~nm. This
elementary configuration is well-known from coupled-mode
theory~\cite{yariv00}. When the gap distance $d$ is reduced the
degenerate modes propagating in the uncoupled waveguides are split
into symmetric and antisymmetric modes.  New propagation
constants $\beta'_s$ and $\beta'_{as}$, respectively are thus
characterizing the symmetric mode and the antisymmetric mode, respectively, and they critically depend on $d$. A
beating of these two modes can be observed in leakage radiation
microscopy~\cite{chen09,colas09} where a mode propagating in one
waveguide can be totally transferred to the second after a coupling
distance $L_c$~\cite{Holmgaard09JLT}. Figures~\ref{fp_Fig4}(b)-(g)
qualitatively show the evolution of the beating pattern and the
coupling distance $L_c$ for decreasing separation distances $d$.
More interesting are the corresponding wave-vector distributions
depicted in the series of FP images in Figs.~\ref{fp_Fig4}(h) to
(m). When the DLSPPWs are weakly coupled ($d$=670~nm), the Fourier
content of Fig.~\ref{fp_Fig4}(h) strongly resembles that of a
single DLSPP mode already shown in Fig.~\ref{fp_Fig3}(b). When $d$ is reduced, a
clear splitting of the modes is observed indicative of a strong
interaction between the waveguides. The parity of the modes can be
readily determined from \textit{e.g.} Fig.~\ref{fp_Fig4}(m). The
asymmetric mode has an odd parity with two maxima centered on each
side of $k_y/k_o$ axis.

\begin{figure}
\includegraphics[width=8.6cm]{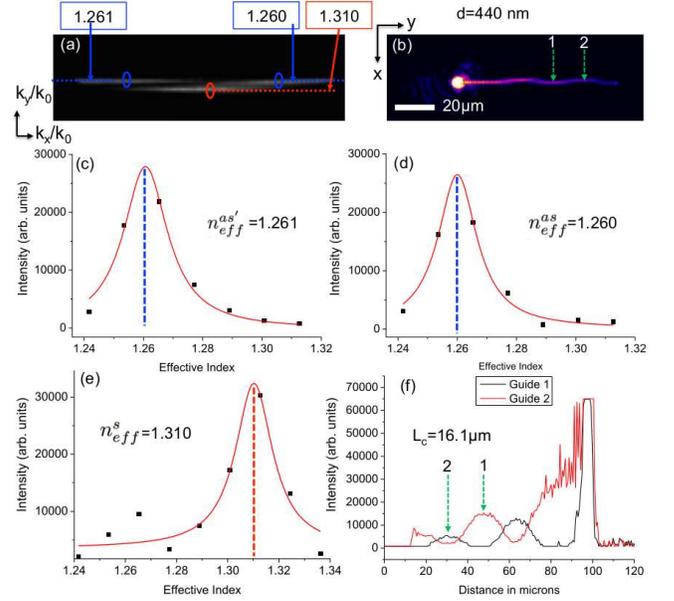}
\caption{(a) and (b) are the respective leakage radiation Fourier
and image planes obtained from a linear DLSPPW coupler with
$d$=440~nm. (c) and (d) are Lorentzian fits of the asymmetric mode
at the location marked by the circle in (a). (e) is a Lorentzian fit
of the symmetric mode. (f) Longitudinal intensity cross sections
taken along the two coupled waveguides in (b) showing the energy
transfer from one DLSPPW to the other defining the coupling length
$L_c$.} \label{fp_Fig5}
\end{figure}

Figure~\ref{fp_Fig5} illustrates the benefit of performing momentum-space spectroscopy described here. Figures~\ref{fp_Fig5}(a)
and (b) are leakage radiation images recorded at the conjugated
Fourier plane and image plane for a linear coupler with $d$=440~nm,
respectively. The effective indices of the symmetric $n_{eff}^{s}$
and antisymmetric $n_{eff}^{as}$ modes are evaluated by Lorentzian
fits of crosscuts of the momentum distribution along the $k_y/k_o$ axis marked by the circles in
Fig.~\ref{fp_Fig5}(a). $n_{eff}^{as}$ was evaluated at two different
wave-vector positions with respect to the $k_y/k_o$ axis, labeled as $n_{eff}^{as}$ and
$n_{eff}^{as^\prime}$. The fits to
the data are represented in Figs.~\ref{fp_Fig5}(c) and (d) for the
antisymmetric mode leading to $n_{eff}^{as}\simeq
n_{eff}^{as^\prime}$=1.260$\pm$0.006. The effective
index of the symmetric mode is measured at
$n_{eff}^s$=1.310$\pm$0.006 (Fig.~\ref{fp_Fig5}(e)).

The coupling length $L_c$ can now be evaluated using the following
relation~\cite{Holmgaard09JLT}
\begin{equation}
L_{c}=\frac{\pi}{\vert \beta'_s - \beta'_{as} \vert},
\end{equation}
where $\beta'_{s}=k_o\times
n_{eff}^{s}$ and $\beta'_{as}=k_o\times n_{eff}^{as}$ . Then
\begin{equation}
L_{c}=\frac{\lambda_0}{2 \vert
n_{eff}^{s}-n_{eff}^{as}\vert}=15.2\rm{\mu m}.
\label{Lc}\end{equation} This procedure was repeated for the
different DLSPPW separation distances $d$. The extracted values of
the split modes and coupling length $L_c$ are reported in
Table~\ref{Gvar}. $L_{c}$ was estimated taking into account an
average value of $n_{eff}^{as}$ and $n_{eff}^{as^\prime}$. For
comparison purposes, $L_c$ was also determined by analyzing the
beating pattern recorded in the image plane (Fig.~\ref{fp_Fig5}(b)).
A longitudinal cross-section of the leakage intensity taken along
each DLSPPWs is shown in Fig.~\ref{fp_Fig5}(f) for $d$=440~nm. The
transfer of the mode between the two waveguides is obtained after
$L_c\sim$16~$\mu$m. Within our pixel resolution, this value compares
well with the $L_c$ determined by the analysis of the degeneracy
lift of the effective indices by Eq.~\ref{Lc}. The advantage of a
momentum-space spectroscopy is that, unlike near-field
measurement~\cite{steinberger07, Holmgaard09JLT}, the effective
indices can be directly measured and mode symmetry
visualized~\cite{grandidier08,Frisbie10}. $L_c$
inferred from direct plane analysis only provides the difference
between the two propagation constants $\vert \beta'_s -\beta'_{as}
\vert$ without discriminating the symmetric mode from the
antisymmetric one.

\begin{table}

\begin{ruledtabular}

    \begin{tabular}{cccccc}

     $d$ [nm] & $n_{eff}^{as}$ & $n_{eff}^{as^\prime}$ & $n_{eff}^{s}$ & $L_c$ [$\mu$m](FP) & $L_c$ [$\mu$m](IP)   \\
\hline
    090& 1.230 & 1.234 & 1.403 & 4.4 & 4.3 \\
    240& 1.241 & 1.244 & 1.343 & 7.5  & 7.7 \\
    350& 1.252 & 1.254 & 1.316 & 12.0 & 11.8 \\
    440& 1.260 & 1.261 & 1.310 & 15.2  & 16.1 \\
    600& 1.262 & 1.263 & 1.299 & 20.6 & 21.1 \\
    670& 1.271 & 1.271 & 1.294 & 32.8 & 31.4 \\
\end{tabular}

\end{ruledtabular}
\caption{\label{tab1} Measured values of the effectives indices of
the asymmetric and symmetric modes and estimated coupling length
$L_c$ for linear DLSPPW couplers with different edge-to-edge
separation $d$. $L_c$ values were estimated independently by
momentum-space spectroscopy and direct-plane analysis.} \label{Gvar}
\end{table}

\begin{figure}
\includegraphics[width=8.6cm]{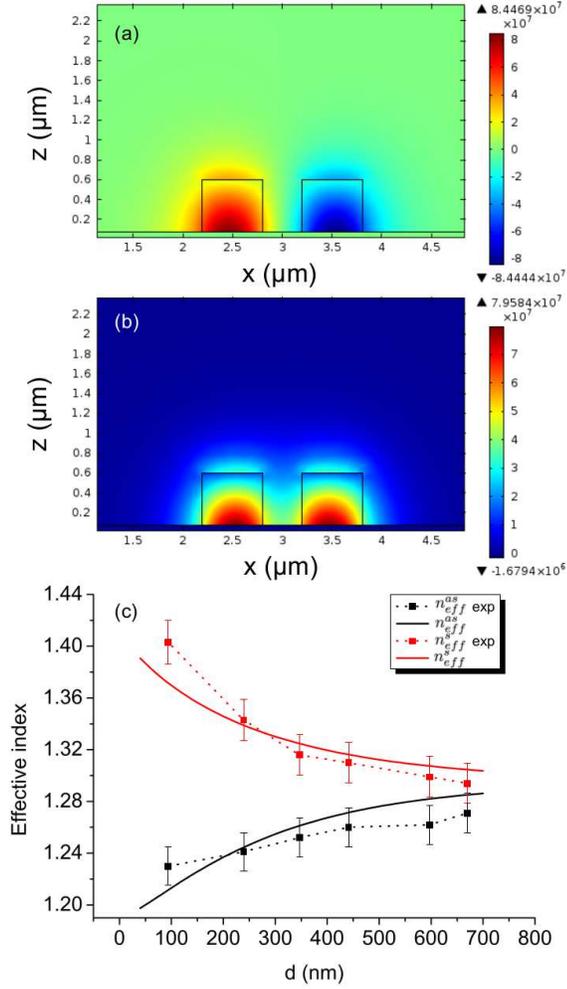}
\caption{(a) and (b) are the electric field profiles of the
asymmetric and symmetric modes propagating in two coupled DLSPPWs
separated by $d$=400~nm. (c) Comparison between the experimental
splitting and the calculated values for various
coupling distances $d$.} \label{fp_Fig6}
\end{figure}

To confirm these experimental results, we numerically simulated
linear DLSPPW couplers with the commercial Finite-Element mode solver
COMSOL. The optical index of the PMMA waveguide is $n_{pmma}=1.535$
and that of the gold layer is $n_{gold}= 0.536 + i 9.5681$ at
$\lambda$=1510~nm~\cite{palik}. The transversal electric field
distribution of the asymmetric and symmetric mode are shown
Figs.~\ref{fp_Fig6} (a) and (b) for $d$=400~nm.
Figure~\ref{fp_Fig6}(c) provides a comparison of the evolution of
the degeneracy lift with separation distance $d$. The measured data
quantitatively reproduce the response of the simulated device.

\section{Momentum-space spectroscopy of curved DLSPPWs}

\subsection{\label{sec:level2} Experimental images}

\begin{figure}
\includegraphics[width=8.6cm]{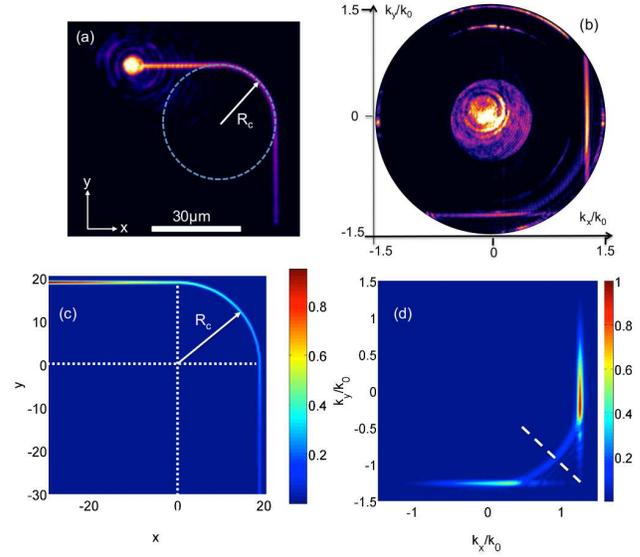}
\caption{(a) and (b) Leakage radiation image of a bent waveguide
with $R_c$=19~$\mu$m and its corresponding wave-vector distribution,
respectively. The Fourier content of the curved section is appearing
as an arc linking the two $k_x$ and $k_y$ lines. (c) and (d)
Computed intensity of the mode propagating in the routing element
and its corresponding wave-vector distribution, respectively.}
\label{fp_Fig7}
\end{figure}

In this section, we investigate the momentum distribution of another
well-known basic routing element: 90$^{\circ}$ curved waveguides.
This DLSPPW structure has been extensively studied by various groups
and the effect of bend radius on the overall losses is well
understood~\cite{Marcatili,krasavin08,Dikken:10,Song:10,Yang:12}. By
performing momentum-space spectroscopy of the supported mode, we
visualize and analyze the wave-vector content of the bend section for this routing device. We demonstrate the limitations of momentum-space spectroscopy to extract modal properties of this elementary building block.
 The curved DLSPPW is composed of two $L$=30~$\mu$m long
straight waveguides with $w$=600~nm. These two input and output
waveguides are linked by a circular $90^{\circ}$ bend section of
radius $R_c$. Bent DLSPPWs with $R_c$ ranging from 5~$\mu$m to
19~$\mu$m were fabricated. The leakage intensity distribution image
of a $R_c$=19~$\mu$m curved waveguide is shown
Fig.~\ref{fp_Fig7}(a). The corresponding wave-vector distribution is
depicted in Fig~\ref{fp_Fig7}(b). The two straight lines at $k_x$
and $k_y$ constant are related to the mode propagating along the $y$
and $x$-oriented waveguides, respectively. The line at
$k_x$=constant is the input waveguide. This Fourier plane shows
additional signatures such as the illumination wave-vector span
(central disk) and planar plasmon modes supported by the Au/air or
Au/PMMA taper interfaces. Of particular interest here is the Fourier
signature of the bent section of the waveguide visible as an arc of
circle linking the two $k_x$ and $k_y$ lines.

\subsection{\label{sec:level3}  Fourier plane model}

\begin{figure*}
\includegraphics[width=\textwidth]{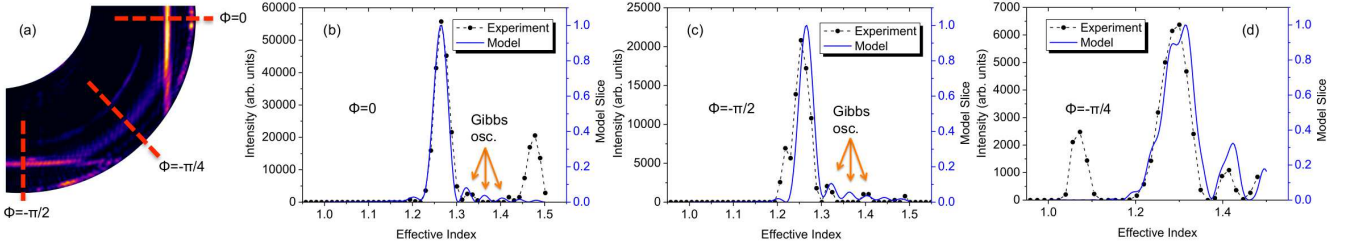}
\caption{(a) Experimental momentum-space image of a $R_c=19~\mu$m
bend waveguide. The dashed lines indicate the position of the
momentum profile in the following graphs. (b) (c) and (d) show the
comparison of the calculated and experimental wave-vector
distributions along the $k_x$, $k_y$ and -$\pi/4$ axis,
respectively.} \label{fp_Fig8}
\end{figure*}
Instead of full numerical simulations, we propose in the following a simple
analytical model that explains the main features of the measured Fourier plane images.
To this aim, we approximated the mode that propagates in the bend structure by a gaussian profile
with a finite propagation length. The characteristics of the gaussian profile
was defined by the experimental data obtained from a straight
waveguide; namely $n_{eff}$=1.262 and $L_{spp}$=42.2~$\mu$m.
The mode width was fixed at $w_0$=500~nm according to the
analysis of Holmgaard and Bozhevolnyi~\cite{holmgaard07}.

The magnetic field is written as follow:
\begin{itemize}
 \item input straight guide ($-L <x<0$, $y>0$)
\begin{eqnarray}
\label{eqstraight}
&&H_y(x,y)=H_o\exp\left[\frac{-(y-R_c)^2}{w_0^2}\right]\exp [i \beta x]\,, \\
&&=H_o\exp\left[\frac{-(y-R_c)^2}{w_0^2}\right]\exp [i n_{eff}k_o x] \exp [-x/2L_{spp}] \,,
\nonumber
\end{eqnarray}
\item circular portion ($x>0$, $y>0$)
\begin{eqnarray}
\label{eq5}
H_r(r,\theta) &=& H_o\exp[-L/2L_{spp}]\\
&&\exp\left[\frac{-(r-R_c)^2}{w_0^2}\right]\exp[i
k_b R_c \theta] \,,
\nonumber
\end{eqnarray}
where $r=\sqrt{x^2+y^2}$, $\theta=\arctan{(x/y)}$ and $k_b$ is the
complex wave-vector of the mode in the curved section.

 \item output straight guide ($x>0$, $-L <y<0$)
\begin{eqnarray}
\label{eqstraighty}
H_x(x,y)&=&H_o\exp\left[\frac{-(x-R_c)^2}{w_0^2}\right]\exp[-L/2L_{spp}]\\
&&\exp[-R_c\pi/4L_{spp}]\exp [i n_{eff}k_o y] \exp[-y/2L_{spp}]\,.
\nonumber
\end{eqnarray}

\end{itemize}

The momentum representation $\tilde{H}(k_x,k_y)$ was obtained by a
Fourier transform
\begin{eqnarray}
\label{eq:FT}
\tilde{H}(k_x,k_y)=\int_{-L}^{R_c+3\times w_0}\int_{-L}^{R_c+3\times
w_0} \! H(x,y)\,\\ \exp\left[ -i(k_xx+k_yy)\right] \mathrm{d} x
\mathrm{d} y. \nonumber
\end{eqnarray}
where the integration window is truncated to limit the calculation on the mode extension.
Outside this area, the mode profile vanishes so that its contribution to the Fourier transform is negligibly small.

To simulate the curved DLSPPW, we maintained the complex propagation constant in
the curved section equal to that of the straight waveguides
($k_b=\beta$). This assumption remains valid for $R_c>R_l$ where
$R_{l}$ is the limiting radius where bend losses can be neglected. $R_l$ have
been recently numerically estimated for DLSPPW~\cite{Song:10}
confirming an analytical expression historically used for standard
optical waveguides~\cite{Marcatili}
\begin{equation}
R_{l} > \frac{24\pi^2 \vert w_3 \vert^3}{\lambda^2}.
\end{equation}
Here $w_3$ corresponds to the length over which the field outside of
the waveguide decays by $1/e$. With $w_3 \simeq w_0$, $R_{l} \simeq$
13~$\mu$m. When the radius is below $R_{l}$, bending losses are
induced by a radial displacement of the mode profile with respect to
the waveguide axis~\cite{Song:10}. This displacement pushes the
field outside the waveguide leading to a lower phase
velocity and a modification of the effective index~\cite{hunsperger}.

Considering curvature loss as an additional exponential decay is a
good approximation to quantify the total loss induced by the bend
and estimate the transmission of the 90$^\circ$ waveguide. However,
this approximation is not representative of the real shape of the field
along the bend and consequently cannot be used to model Fourier images of the kind displayed in Fig.~\ref{fp_Fig7}(b).

For $R_c> R_l\simeq$13~$\mu$m, we show that the propagation and
momentum-space representations simulated with these basic
assumptions (Figs.~\ref{fp_Fig7}(c) and (d)) reproduces the
experimental observations of Figs.~\ref{fp_Fig7}(a) and (b).
We compare the profile of the calculated
Fourier transform (Fig.~\ref{fp_Fig7}(d)) to the experimental wave-vector distribution (Fig.~\ref{fp_Fig7}(b)) by
extracting momentum profiles at different radial coordinates
indicated in Fig.~\ref{fp_Fig8}(a) for $R_c$=19~$\mu$m.
Figures~\ref{fp_Fig8}(b), (c) and (d) show the calculated and
experimental wave-vector distributions taken along the $k_x$ and
$k_y$ axis and at -$\pi/4$, respectively. The position and width of
the calculated wave-vector profiles are in very good agreement with
the measured data validating thus our preliminary assumptions. Some
extra signatures are visible on the experimental cuts: a peak at
1.49 in Fig.~\ref{fp_Fig8}(b) corresponds to the numerical aperture
of the collection objective, and the contribution of the surface
plasmon excited in the nearby gold film is visible at 1.08 in
Fig.~\ref{fp_Fig8}(d). Noticeable also is the presence of Gibbs
oscillations revealed by this momentum-space spectroscopy (arrows Figs.~\ref{fp_Fig8} (b) and (c)) . The
origin of these oscillations in reciprocal space is discussed below.

\subsection{\label{sec:level4}  Analytical development}

In order to provide analytical expressions and propose a simple physical understanding
of the measured Fourier images, we simplified further the model. We consider each part of the waveguide (straight and bend) independently from each others.
%
The straight parts are approximatively 30~$\mu$m
long, a length smaller than the longitudinal decay of the supported
mode ($L_{spp}$=42~$\mu$m). If we consider only the field in the input
straight waveguide along the $x$ coordinate, the Fourier transform
of the magnetic field written in Eq.~\ref{eqstraight} reads
\begin{equation}
\tilde{H}_y(k_x)=H_o\int_{-L}^{0}\! \exp[i (\beta-k_x)x]\,
\mathrm{d} x \,,
\end{equation}
where we neglected the $y$-gaussian profile for the sake of clarity but this could be easily considered.
Then the intensity in the Fourier plane along $k_x$ direction
writes
\begin{equation}
\vert \tilde{H}_y(k_x)
\vert^2=|H_o|^2\frac{1-2e^{-\beta^{\prime\prime}L}\cos((\beta^{\prime}-k_x)L)+e^{-2\beta^{\prime\prime}L}}{(\beta^{\prime}-k_x)^2+\beta^{\prime\prime
2}} \label{eq9}
\end{equation}

\begin{figure}
\includegraphics[width=8.6cm]{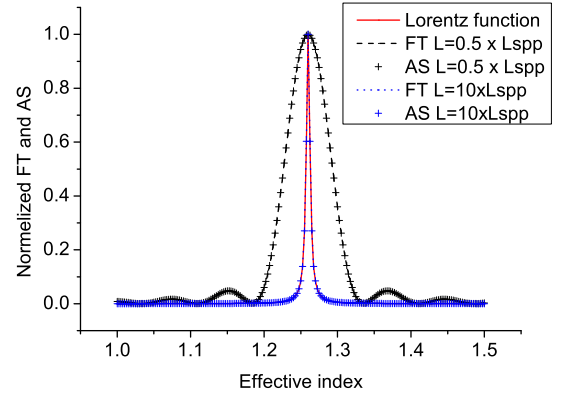}
\caption{Effect of the DLSPPW length on Fourier plane calculations. Dashed lines: calculated Fourier plane (FT) cross section for $L=0.5\times L_{spp}$ and $L=10\times L_{spp}$. Crosses: analytical
solutions (AS) calculated from Eq.~\ref{eq9} with the same integration
boundaries. Red solid line: calculated Lorentzian
profile of the wave-vector distribution of a plasmon mode with
a propagation length of 40~$\mu$m. When $L<L_{spp}$ Gibbs oscillations and widening of the wave-vector content occur.} \label{fp_Fig9}
\end{figure}

If $L \rightarrow \infty$
then $\vert \tilde{H}_y(k_x) \vert^2
\rightarrow 1/[(\beta^{\prime}-k_x)^2+\beta^{\prime\prime 2}]$ and follows a Lorentzian profile, as expected.
If the straight part length $L$ is smaller than $L_{spp}$,
the resulting Fourier transform is not simply defined by a
Lorentzian function. Because of the finite integration limit, extra oscillations are becoming visible (Gibbs oscillations) as illustrated in Figs.~\ref{fp_Fig8}(b) and (c).
For long waveguide
(as in Fig.~\ref{fp_Fig3}(a) where $L>L_{spp}$), no oscillation of
the wave-vector distribution is observed. Figure~\ref{fp_Fig9} shows
the impact of the integration boundaries $L$ on the momentum space
representation of a single straight waveguide for a fixed
propagation length of 40~$\mu$m. There is a perfect agreement
between the numerically calculated Fourier transform and its analytical solution given by
Eq.~\ref{eq9}. For integration boundaries greater than the
propagation length the two calculations match with the Lorentz
function generally used.

Since the momentum space representation of a straight waveguide can
be expressed for any integration boundaries, we now look for the
expression of a curved waveguide alone. On
Fig.~\ref{fp_Fig8} (d), the $-\pi/4$ profile shows the experimental
momentum space of a $90^{\circ}$ curved waveguide linked by an input
and an output waveguide. We approximate in the following a solution
considering a fully symmetric configuration consisting of a lossless
circular waveguide. Losses are omitted to maintain the radial
symmetry valid between [0;2$\pi$]. The Fourier transform in a $360^{\circ}$
bend can be expressed in polar coordinates as

\begin{equation}
\tilde{H}(k_r,\phi)=\int_0^{\infty}\int_0^{2\pi}  \! H_r(r,\theta)
e^{-2i\pi k_r r \cos(\theta-\phi)} r\, \mathrm{d} r  \mathrm{d}
\theta.\label{eq10}
\end{equation}

$H_r(r,\theta)$ is given by Eq.~\ref{eq5}. Since this function is
periodic with respect to $\theta$ between [0;2$\pi$], we can expand
$H_r(r,\theta)$ in a Fourier series:

\begin{equation}
H_r(r,\theta )=\sum _{ n=-\infty  }^{ +\infty  }{ H_{ n }(r) } e^{
in\theta },
\end{equation}
where the $n^{{\rm th}}$ harmonic is
\begin{eqnarray}
 H _{n}( r )&=&\frac{1}{2\pi}\int_{-\pi}^{\pi} {H_r(r,\theta) } e^{- in\theta } \mathrm{d}
 \theta \label{eq12}\\
 &=&e^{\left[\frac{-(r-R_c)}{w_0^2}\right]} {\rm sinc}(k_b R_c
 \pi-n\pi). \nonumber
\end{eqnarray}
The Fourier transform of the field (Eq.~\ref{eq10}) can now be
written as follow
\begin{equation}
\tilde { H }_{n} ({ k }_{ r },\phi )=\sum _{ n=-\infty  }^{ +\infty  }{
(-i)^{ n } } e^{ i\phi n }2\pi \int _{ 0 }^{ \infty  }{ rH_{ n
}(r)J_{ n }(2\pi k_{ r }r) } dr, \label{eq13}\\
\end{equation}
\begin{equation}
\tilde { H }_{n} ({ k }_{ r },\phi )=\sum _{ n=-\infty  }^{ +\infty  }{h_n}{f_n({ k }_{ r },\phi)},
\end{equation}
with
\begin{equation}
{h_n}={\rm sinc}(k_b R_c\pi-n\pi), \label{eq14}
\end{equation}
and
\begin{equation}
f_n({ k }_{ r },\phi)={(-i)^{ n } } e^{ i\phi n }2\pi \int _{ 0 }^{ \infty  }{ re^{\left[\frac{-(r-R_c)}{w_0^2}\right]}J_{ n }(2\pi k_{ r }r) } dr.\nonumber
\end{equation}

$J_n(2\pi k_r r)$ is the Bessel function of the first kind of
$n^{{\rm th}}$ order.
Further simplification can be achieved by extracting the main
$n^{{\rm th}}$ harmonic $h_n$ of the series. It is deduced from Eq. \ref{eq14} and is such that

\begin{eqnarray}
k_b R_c \pi - n \pi = 0 <=> n=k_b R_c.
\end{eqnarray}
Therefore the Fourier series is dominated by the harmonic integer that is the closest to the
product $k_bR_c$.
\begin{figure}
\includegraphics[width=8.6cm]{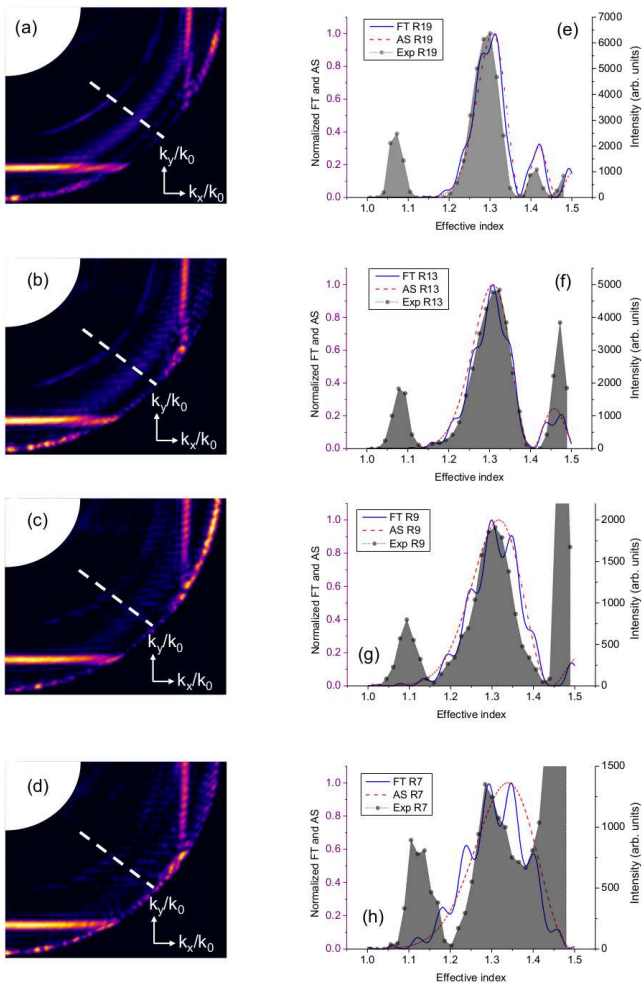}
\caption{(a-d) Selected region of interest of the experimental
momentum-space images for curved waveguides with $R_c$=19~$\mu$m,
13~$\mu$m, 9~$\mu$m, and 7~$\mu$m, respectively. (e) to (h)
Experimental profiles (shaded areas) of the wave-vector distribution
taken along the dashed line in (a) to (d). The solid blue lines are
the calculated profiles using Eq.~\ref{eq:FT} already reported in Fig.~\ref{fp_Fig8}. The red dashed lines
show the analytical solutions obtained with the $n^{{\rm th}}$ order
rendering the best agreement with the data. $n$=100, 68, 43, and
37, respectively.} \label{fp_Fig10}
\end{figure}

Figure~\ref{fp_Fig10} illustrates a comparison between the
experimental data extracted from momentum-space spectroscopy, the
Fourier calculations derived from Eq.~\ref{eq:FT} already used in Figure~\ref{fp_Fig8}, and the analytical
approximation discussed above. The experimental Fourier planes of
Figs.~\ref{fp_Fig10} (a) to (d) only show the region of interest
for four different radii. The corresponding wave-vector profiles
extracted and calculated at $-\pi$/4 (dashed lines in figure~\ref{fp_Fig7}(d)) are plotted in
the graphs of Figs.~\ref{fp_Fig10} (e) to (h). For radius $R_c>R_l$,
the Fourier calculations and the analytical solutions are in good
agreement with the experimental data.
For large radii the signature of the bend in momentum space is
defined only by the phase difference $k_b(2\pi R_c)=n2\pi$.
This corresponds to resonance condition of a ring resonator.

Experimentally, the periodic condition on $\theta$ is not respected
since the structure considered is formed by an arc of circle.
Nonetheless, using the $n=100^{{\rm th}}$ order of the Bessel
function in Eq.~ \ref{eq13} for $R_c$=19~$\mu$m and the $68^{{\rm
th}}$ order for $R_c$=13~$\mu$m  the experimental data and the
calculated Fourier transform can be well reproduced
(Figs.~\ref{fp_Fig10} (e) and (f)). For radius $R_c<R_l$ , the
calculated profiles deduced from Eq.~\ref{eq:FT} and from the
$n=47^{{\rm th}}$ and $n=37^{{\rm th}}$ orders for $R_c$=9~$\mu$m and
$R_c$=7~$\mu$m, respectively are deviating from the experimental
cross-cuts (Figs.~\ref{fp_Fig10} (g) and (h)). This disagreement is
expected since none of the two models (Fourier transform and
analytical approximation) are including bending loss.

\section{conclusion}

By using dual-plane leakage radiation microscopy we have fully
quantified the key parameters characterizing two important
dielectric-loaded surface plasmon polariton routing devices: linear
couplers and 90$^{\circ}$ curved waveguides. We unambiguously
demonstrated the added-value of performing momentum-space
spectroscopy. The degeneracy lift for strongly coupled waveguides
and the symmetry of the split modes can be directly visualized and
quantified. The wave-vector distribution associated to the curved
section of the waveguide was also readily observed. We developed a
numerical and an analytical analyzis to understand the experimental
momentum distribution. We found that for large radii (vanishing
bending loss), we can link the plasmon signature in Fourier space with the geometrical and modal properties of the bend structure. The radial dependence of the wave-vector distribution is governed by the phase difference $k_b(2\pi R_c)$. For smaller radii of curvature the bend loss need to be accounted for by developing an approach including realistic field shape. 

\section{Acknowledgments}
This work was funded by the European FP7 research program PLATON,
Contract Number 249135 and the European Research Council grant agreement number 306772, the regional council of Burgundy under
the program PARI SMT3 and the Labex ACTION. S. Lach\`eze is acknowledged
for participating at the early stage of this work, with the support of Burgundy Region and CEA Leti Carnot funding.

\end{document}